\documentclass[aps,prl,twocolumn,superscriptaddress,showpacs]{revtex4-1}

\usepackage{graphicx}
\usepackage{dcolumn}
\usepackage{bm}
\usepackage{natbib}

\begin{document}


\title{Single microwave-photon detector using an artificial $\Lambda$-type three-level system}


\author{Kunihiro Inomata$^{\dagger}$}
\email[Corresponding author. E-mail: \\]{kunihiro.inomata@riken.jp}
\affiliation{RIKEN Center for Emergent Matter Science (CEMS), Wako, Saitama 351-0198, Japan}
\author{Zhirong Lin}
\thanks{These authors contributed equally to this work.}
\affiliation{RIKEN Center for Emergent Matter Science (CEMS), Wako, Saitama 351-0198, Japan}
\author{Kazuki Koshino}
\affiliation{College of Liberal Arts and Sciences, Tokyo Medical and Dental University, Ichikawa, Chiba 272-0827, Japan}
\author{William D. Oliver}
\affiliation{MIT Lincoln Laboratory, Lexington, Massachusetts 02420, USA}
\affiliation{Departent of Physics, Massachusetts Institute of Technology, Cambridge, Massachusetts 02139, USA}
\author{Jaw-Shen Tsai}
\affiliation{RIKEN Center for Emergent Matter Science (CEMS), Wako, Saitama 351-0198, Japan}
\affiliation{Department of Physics, Tokyo University of Science, Shinjuku-ku, Tokyo 162-8601, Japan}
\author{Tsuyoshi Yamamoto}
\affiliation{NEC Smart Energy Research Laboratories, Tsukuba, Ibaraki 305-8501, Japan}
\author{Yasunobu Nakamura}
\affiliation{RIKEN Center for Emergent Matter Science (CEMS), Wako, Saitama 351-0198, Japan}
\affiliation{Research Center for Advanced Science and Technology (RCAST), The University of Tokyo, Meguro-ku, Tokyo 153-8904, Japan}

\begin{abstract}
Single photon detection is a requisite technique in quantum-optics experiments in both the optical and the microwave domains.
However, the energy of microwave quanta are four to five orders of magnitude less than their optical counterpart, making the efficient detection of single microwave photons extremely challenging.
Here, we demonstrate the detection of a single microwave photon propagating through a waveguide.
The detector is implemented with an ``impedance-matched" artificial $\Lambda$ system comprising the dressed states of a driven superconducting qubit coupled to a microwave resonator.
We attain a single-photon detection efficiency of $0.66 \pm 0.06$ with a reset time of $\sim 400$~ns.
This detector can be exploited for various applications in quantum sensing, quantum communication and quantum information processing.
\end{abstract}


\maketitle
Single-photon detection is essential to many quantum-optics experiments, enabling photon counting and its statistical and correlational analyses~\cite{Hadfield09}.
It is also an indispensable tool in many protocols for quantum communication and quantum information processing~\cite{Gisin02,Knill01,Brien07,Aaronson11}.
In the optical domain, various kinds of single-photon detectors are commercially available and commonly used~\cite{Hadfield09,Eisaman11}.
However, despite the latest developments in nearly-quantum-limited amplification~\cite{Bergeal10,Macklin15} and homodyne measurement for extracting microwave photon statistics~\cite{Lang13}, the detection of a single microwave photon in an itinerant mode remains a challenging task due to its correspondingly small energy.
Meanwhile, the demand for such detectors is rapidly increasing, driven by applications involving both microwave and hybrid optical-microwave quantum systems.

In this report we demonstrate an efficient and practical single-microwave-photon detector based on the deterministic switching in an artificial $\Lambda$-type three-level system implemented using the dressed states of a driven superconducting quantum circuit.
The detector operates in a time-gated mode and features a high quantum efficiency $0.66 \pm 0.06$, a low dark-count probability $0.014 \pm 0.001$, a bandwidth $\sim 2\pi \times 16$~MHz, and a fast reset time $\sim 400$~ns. It can be readily integrated with other components for microwave quantum optics.

Our detection scheme carries several advantages compared with previous proposals.
It uses coherent quantum dynamics, which minimizes energy dissipation upon detection and allows for rapid resetting with a resonant drive, in contrast to schemes that involve switching from metastable states of a current-biased Josephson junction into the finite voltage state~\cite{Romero09,Peropadre11,Chen11}.
Moreover, our detection scheme does not require any temporal shaping of the input photons, nor precise time-dependent control of system parameters adapted to the temporal mode of the input photons, in contrast to recent photon-capturing experiments~\cite{Yin13,Palomaki13,Wenner14}.
It also achieves a high efficiency without cascading many devices~\cite{Romero09,Sathy14}.

The operating principle of the detector fully employs the elegance of waveguide quantum electrodynamics, which has recently attracted significant attention in various contexts surrounding photonic quantum information processing~\cite{Duan04,Chang07,Witthaut10,Zheng13}.
When electromagnetic waves are confined and propagate in an one-dimensional (1D) mode, their interaction with a quantum emitter/scatterer is substantially simplified and enhanced compared with three-dimensional cases.
These advantages result from the natural spatial-mode matching of the emitter/scatterer with a 1D mode and its resulting enhancement of quantum interference effects.
Remarkable examples are the perfect extinction of microwave transmission for an artificial atom coupled to a 1D transmission line~\cite{Oleg10,Hoi11}, the photon-mediated interaction between two remote atoms coupled to a 1D transmission line~\cite{vanLoo13}, and the perfect absorption --- and thus ``impedance matching" --- of a $\Lambda$-type three-level system terminating a 1D transmission line~\cite{Koshino13,Inomata14}.
In the latter system, the incident photon deterministically induces a Raman transition which switches the state of the $\Lambda$ system~\cite{Pinotsi08,Koshino13}.
This effect has recently been demonstrated in both the microwave and optical domains~\cite{Inomata14,Shomroni14}, indicating its potential for photon detection~\cite{Koshino15} as well as for implementing deterministic entangling gates with photonic qubits~\cite{Koshino10}.
\begin{figure*}
\includegraphics[width=17cm]{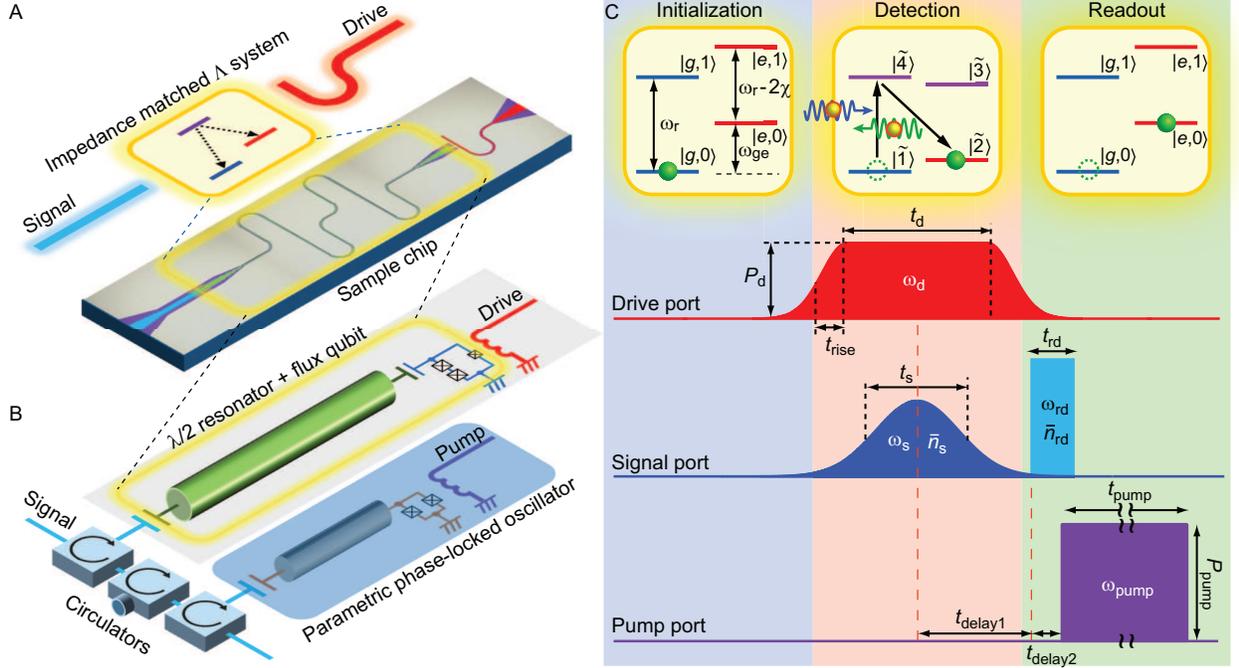}
\caption{Experimental setup and pulse sequence.
(A)~Image of the sample chip containing a flux qubit and a superconducting microwave resonator coupled capacitively and operated in the dispersive regime. For certain proper conditions of the qubit drive, the coupled system functions as an impedance-matched $\Lambda$-type three-level system.
(B)~Schematic of the itinerant microwave-photon detector consisting of the coupled system and connected to a parametric phase-locked oscillator (PPLO) via three circulators in series. The circuit has three input ports: signal, qubit drive, and pump for the PPLO.
(C)~Energy-level diagram of the coupled system and the pulse sequence for single-photon detection. The system is first prepared in the ground state. During the detection stage, we concurrently apply the drive and signal pulses. The drive is parameterized to fulfill the impedance-matched condition such that a signal photon (blue arrow) induces a deterministic Raman transition.
A down-converted photon (green arrow) is emitted in the process and discarded.
In the readout stage, we detect the qubit excited state nondestructively by sending a qubit readout pulse.
The qubit-state-dependent phase shift in the reflected pulse is discriminated by the PPLO.
}
\end{figure*}

Our device consists of a superconducting flux qubit capacitively and dispersively coupled to a microwave resonator (Fig.~1B)~\cite{Inomata12}.
With a proper choice of the qubit drive frequency $\omega_{\rm d}$ and power $P_{\rm d}$, the system functions as an impedance-matched $\Lambda$ system with identical radiative decay rates from its upper state to its two lower states (Fig.~1A)~\cite{Koshino13,Inomata14}.
The qubit-resonator coupled system is connected to a parametric phase-locked oscillator (PPLO), which enables fast and non-destructive qubit readout~\cite{Lin14}.

Figure~1C shows the level structure of the qubit-resonator system and the protocol for the single photon detection.
We label the energy levels $\vert q, n \rangle$ and their eigenfrequencies $\omega_{\vert q, n \rangle}$, where $q=\{g, e\}$ and $n = \{0, 1, \cdots\}$ respectively denote the qubit state and the photon number in the resonator.
In the dispersive coupling regime, the qubit-resonator interaction renormalizes the eigenfrequencies to yield $\omega_{\vert g, n \rangle}=n\omega_{\rm r}$ and $\omega_{\vert e, n \rangle}=\omega_{\rm ge} + n(\omega_{\rm r}-2\chi)$, where $\omega_{\rm ge}$ and $\omega_{\rm r}$ are the renormalized frequencies of the qubit and the resonator, respectively, and $\chi$ is the dispersive frequency shift of the resonator due to its interaction with the qubit.
Only the lowest four levels with $n=0$ or $1$ are relevant here.

We prepare the system in its ground state $\vert g, 0\rangle$ (Fig.~1C, Initialization) and apply a drive pulse to the qubit (Fig.~1C, Detection).
In a frame rotating at $\omega_{\rm d}$, the level structure becomes nested, i.e., $\omega_{\vert g, 0 \rangle} < \omega_{\vert e, 0 \rangle} < \omega_{\vert e, 1 \rangle} < \omega_{\vert g, 1 \rangle}$, for $\omega_{\rm d}$ in the range $\omega_{\rm ge}-2\chi < \omega_{\rm d} < \omega_{\rm ge}$.
On the plateau of the drive pulse ($P_{\rm d}>0$), the lower-two levels $\vert g, 0 \rangle$ and $\vert e, 0 \rangle$ (higher-two levels $\vert g, 1 \rangle$ and $\vert e, 1 \rangle$) hybridize to form dressed states $\vert \tilde{1} \rangle$ and $\vert \tilde{2} \rangle$ ($\vert \tilde{3} \rangle$ and $\vert \tilde{4} \rangle$).
Under a proper choice of $P_{\rm d}$, the two radiative decay rates from $\vert \tilde{4} \rangle$ (or $\vert \tilde{3} \rangle$) to the lowest-two levels become identical.
Thus, an impedance-matched $\Lambda$ system comprising $\vert \tilde{1} \rangle$, $\vert \tilde{2} \rangle$, and $\vert \tilde{4} \rangle$ (alternatively, $\vert \tilde{1} \rangle$, $\vert \tilde{2} \rangle$, and $\vert \tilde{3} \rangle$) is realized.
An incident single microwave photon (Gaussian envelope, length $t_{\rm s}$), synchronously applied with the drive pulse through the signal port and in resonance with the $\vert \tilde{1} \rangle \rightarrow \vert \tilde{4} \rangle$ transition, deterministically induces a Raman transition, $\vert \tilde{1} \rangle \rightarrow \vert \tilde{4} \rangle \rightarrow \vert \tilde{2} \rangle$, and is down-converted to a photon at the $\vert \tilde{4} \rangle \rightarrow \vert \tilde{2} \rangle$ transition frequency. 
This process is necessarily accompanied by an excitation of the qubit~\cite{Koshino13,Inomata14}.
\begin{figure}[h]
\includegraphics[width=8.5cm]{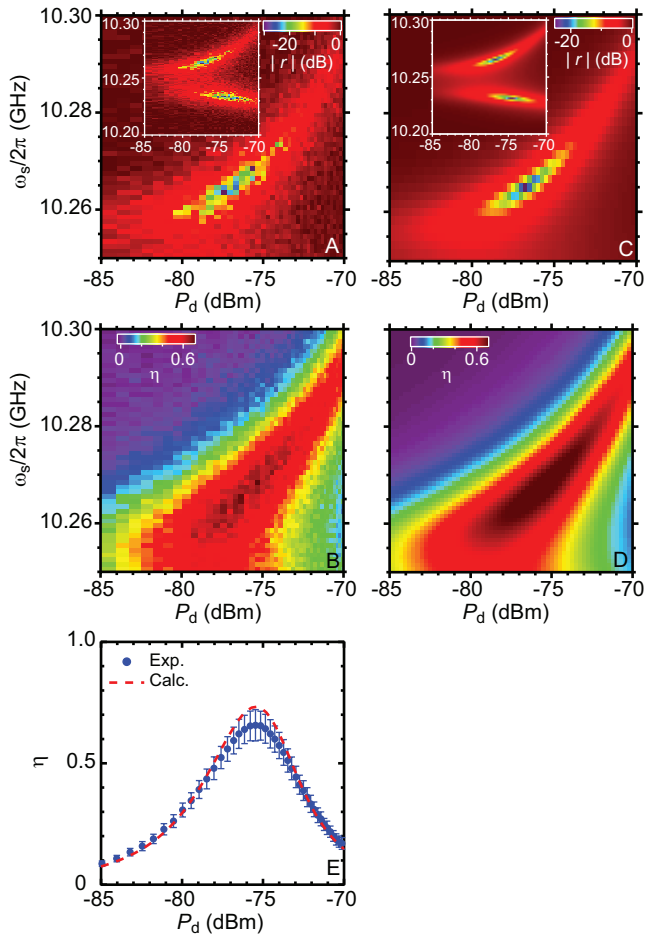}
\caption{Impedance matching and itinerant microwave-photon detection.
(A)~Amplitude of the reflection coefficient $\vert r \vert$ of the input signal pulse with mean photon number $\bar{n}_{\rm s} \sim 0.1$ as a function of the qubit drive power $P_{\rm d}$ and the signal frequency $\omega_{\rm s}$.
The PPLO is not activated during this measurement.
The impedance-matched point is resolved (dark-blue region), where the input microwave photon is absorbed almost completely.
In the inset, we also observe another dip in $\vert r \vert$, corresponding to the Raman transition of $\vert \tilde{1} \rangle \rightarrow \vert \tilde{3} \rangle \rightarrow \vert \tilde{2} \rangle $.
Microwave power levels stated in this article are referred to the value at the corresponding ports on the sample chip.
(B)~Detection efficiency $\eta$ of an itinerant microwave photon. The efficiency hits the maximum at the impedance-matched point, where the Raman transition of $\vert \tilde{1} \rangle \rightarrow \vert \tilde{4} \rangle \rightarrow \vert \tilde{2} \rangle$ takes place.
(C)~and (D)~Theoretical predictions corresponding to A and B.
(E)~Cross-sections of B (blue dots) and D (red dashed line) at $\omega_s/2\pi=10.268~{\rm GHz}$. The error bars are due to the uncertainty in the input power calibration.
}
\end{figure}

To detect the photon, we adiabatically switch off the qubit drive and dispersively read out the qubit state (Fig.~1C, Readout).
We apply a readout pulse with the frequency $\omega_{\rm rd} = \omega_{\rm r} - 2\chi =\omega_{\vert e, 1 \rangle} - \omega_{\vert e, 0\rangle}$ through the signal port, which, 
upon reflection at the resonator, acquires a qubit-state-dependent phase shift of $0$ or $\pi$.
This phase shift is detected by the PPLO with high fidelity: in the present setup, the readout fidelity of the qubit is $\sim 0.9$, which is primarily limited by qubit relaxation prior to readout\cite{Lin14}.

We first determine the operating point where the $\Lambda$ system deterministically absorbs a signal photon.
We simultaneously apply a drive pulse of length $t_{\rm d}=178$~ns and a signal pulse of length $t_{\rm s}=85$~ns, and proceed to measure the reflection coefficient $\vert r \vert$ of the signal pulse as a function of the drive power $P_{\rm d}$  and the signal frequency $\omega_{\rm s}$ (Fig.~2A).
The signal pulse is in a weak coherent state with mean photon number $\bar{n}_{\rm s} \sim 0.1$.
A pronounced dip with a depth of $<$$-25$~dB is observed in $\vert r \vert$ at $(P_{\rm d}, \omega_{\rm s}/2\pi)=(-76~{\rm dBm}, 10.268~{\rm GHz})$, in close agreement with theory (Fig.~2C).
The dip indicates a near-perfect absorption condition, i.e., impedance matching, where the reflection of the input microwave photon vanishes due to destructive self-interference.
Correspondingly, a deterministic Raman transition of $\vert \tilde{1} \rangle \rightarrow \vert \tilde{4} \rangle \rightarrow \vert \tilde{2} \rangle$ is induced, and the qubit state is flipped.
\begin{figure}
\includegraphics[width=5.4cm]{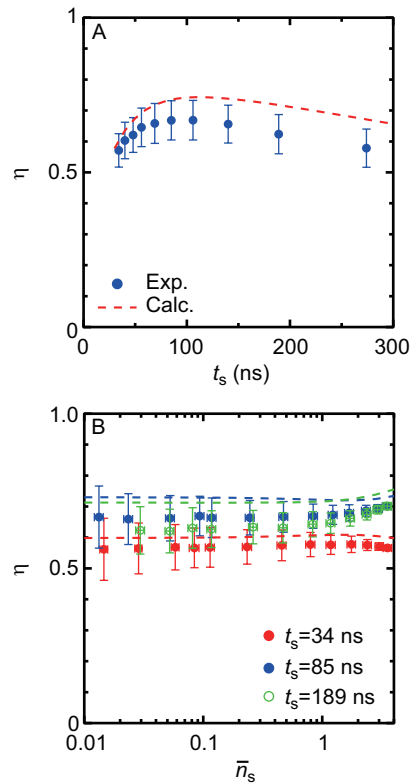}
\caption{Optimization of the efficiency.
(A)~Single photon detection efficiency $\eta$ as a function of the signal pulse length $t_{\rm s}$. The mean photon number $\bar{n}_{\rm s}$ for the weak-coherent signal pulse is $\sim 0.1$.
(B)~$\eta$ as a function of $\bar{n}_{\rm s}$.
Dashed lines indicate theoretical predictions.
}
\end{figure}
\begin{figure*}[]
\includegraphics[width=10.6cm]{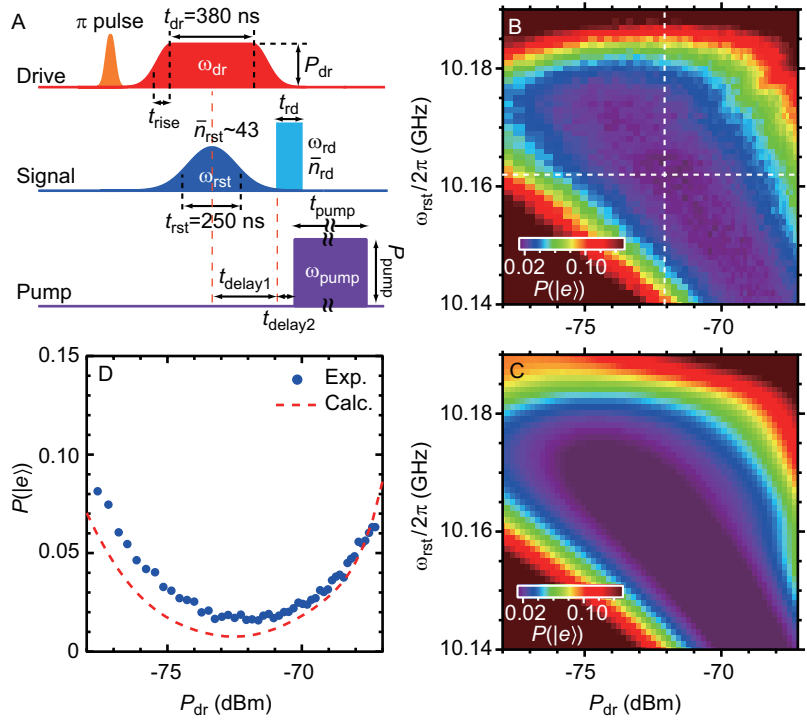}
\caption{Demonstration of the fast reset protocol.
(A)~Pulse sequence used to evaluate the reset efficiency.
The initial $\pi$-pulse mimics a single-photon detection and excites the qubit.
During the reset stage, a drive pulse and a reset pulse with the mean photon number of $\bar{n}_{\rm rst} \sim 43$ are concurrently applied, inducing an inverse Raman transition: $\vert \tilde{2} \rangle \rightarrow \vert \tilde{3} \rangle \rightarrow \vert \tilde{1} \rangle$.
The remaining population in the $\vert e \rangle$ state is then detected.
(B)~Population of the qubit excited state after the reset operation $P(\vert e \rangle)$, as a function of the reset-pulse frequency $\omega_{\rm rst}$ and the drive-pulse power $P_{\rm dr}$.
(C)~Theoretical prediction for B with no free parameters.
(D)~Cross sections of B (blue dots) and C (red dashed line) at $\omega_{\rm rst}/2\pi=10.162$~{\rm GHz}.
}
\end{figure*}

To obtain a `click' corresponding to single-photon detection, we read out the qubit state by using the PPLO immediately after the Raman transition.
Before initiating readout, the drive pulse is turned off to suppress unwanted Raman transitions induced by the readout pulse, e.g., $\vert \tilde{2} \rangle \rightarrow \vert \tilde{3} \rangle \rightarrow \vert \tilde{1} \rangle$.
We repeatedly apply the pulse sequence in Fig.~1C $10^4$ times and evaluate the single-photon detection efficiency $\eta \equiv P(\vert e \rangle)/[1-P(0)]$, where $P(\vert e \rangle)$ and $P(0)$ are the probabilities for the qubit being in the excited state and the signal pulse being in the vacuum state, respectively.
Figure 2B depicts $\eta$ as a function of $P_{\rm d}$ and $\omega_{\rm s}$.
The dark count probability of the detector --- mainly caused by the nonadiabatic qubit excitation due to the drive pulse and the imperfect initialization --- is subtracted when evaluating $\eta$~\cite{Supp_Note}.
We observe that $\eta$ is maximized at the dip position in Fig.~2A in accordance with the impedance-matching condition.
We also confirm that the result agrees with numerical calculations based on the parameters determined independently (Fig.~2D).
The maximum value, $\eta = 0.66 \pm 0.06$, is obtained at $(P_{\rm d}, \omega_{\rm s}/2\pi)=(-75.5~{\rm dBm}, 10.268~{\rm GHz})$  (Fig.~2E)~\cite{Supp_Note}.
The efficiency exceeds 0.5 over a signal-frequency range of $\sim 20$~MHz, which is comparable to the bandwidth of the detector, $\kappa/2\pi \sim 16$~MHz~\cite{Supp_Note}.

In the Fig.~3A, we plot efficiency $\eta$ as a function of the signal pulse length $t_{\rm s}$.
Here, we fix $\omega_{\rm s}$ and $P_{\rm d}$ at the values which maximize $\eta$ in Fig.~2E. 
The drive pulse duration $t_{\rm d}$ is set to be $t_{\rm d}=1.5t_{\rm s} + 50$~ns, which empirically maximizes $\eta$ at each $t_{\rm s}$.
We observe that $\eta$ is a non-monotonic function of $t_{\rm s}$ and attains a maximum at $t_{\rm s} = 85$~ns.
The initial increase of $\eta$ at short $t_{\rm s}$ is due to the narrowing of the signal bandwidth 
resulting in an improved overlap with the detection bandwidth. 
For longer $t_{\rm s}$, the qubit relaxation limits $\eta$~\cite{Koshino15}.
Next, we examine how the photon detector behaves when $\bar{n}_{\rm s}$ in the signal pulse is varied.
Figure 3B shows $\eta$ as a function of $\bar{n}_{\rm s}$ for fixed signal pulse lengths at $t_{\rm s} = 34$, $85$, and $189$~ns.
The detection efficiencies stay constant for $\bar{n}_{\rm s} \lesssim 1$ regardless of the pulse lengths.
This validates the determination of $\eta$ in our measurements using signal pulses in the weak coherent states.
For $\bar{n}_{\rm s} > 1$, $\eta$ slightly depends on $\bar{n}_{\rm s}$ because of the possibility to drive multiple Raman transitions.

After a single-photon detection event, the qubit remains in the excited state until it spontaneously relaxes to the ground state, which leads to a relatively long dead time of the detector.
However, our coherent approach allows us to implement a fast reset protocol (Fig.~4A): in conjunction with the drive pulse that forms the $\Lambda$ system, we apply a relatively strong reset pulse through the signal port which induces an inverse Raman transition, $\vert \tilde{2} \rangle \rightarrow \vert \tilde{3} \rangle \rightarrow \vert \tilde{1} \rangle$.
We optimize the drive-pulse power $P_{\rm dr}$ and the reset-pulse frequency $\omega_{\rm rst}$ such that the resulting qubit excitation probability $P(\vert e \rangle)$ is minimized (Fig.~4B).
At the optimal reset point $(P_{\rm dr}, \omega_{\rm rst}/2\pi) = (-72.1~{\rm dBm}, 10.162~{\rm GHz})$, $P(\vert e \rangle)$ attains a minimum value $0.017 \pm 0.002$, equivalent to the value $0.016 \pm 0.001$ obtained in the absence of the initial $\pi$-pulse used to mimic a photon absorption event.
Without a reset pulse, we obtain $P(\vert e \rangle) = 0.490 \pm 0.010$.
A comparison of the two results indicates that the reset pulse is highly efficient.
Moreover, we confirm that the reset protocol does not affect the succeeding detection efficiency and that the time-gated operation can be repeated in a rate exceeding 1~MHz~\cite{Supp_Note}.

For the moment, the detection efficiency of this detector is limited by the relatively short qubit relaxation time, $T_1 \sim 0.7$~$\mu$s.
Nonetheless, our theoretical work indicates that efficiencies reaching $\sim 0.9$ are readily achievable with only a modest improvement of the qubit lifetime~\cite{Koshino15}.
An extension from the time-gated-mode to the continuous-mode operation is also possible~\cite{Koshino15_2}.\\

This work was partially supported by JSPS KAKENHI (Grant Number 25400417, 26220601, 15K17731), ImPACT Program of Council for Science, and the NICT Commissioned Research.

\bibliography{scifile}

\newpage
\appendix
\section{Supplemental material for\\
``Single microwave-photon detector using an artificial $\Lambda$-type three-level system"}
\subsection{Device}
\begin{figure}
\includegraphics[width=8.5cm]{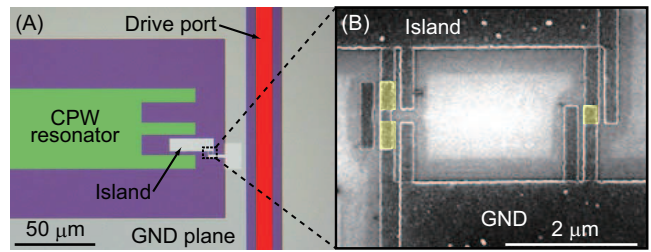}
\centering
\caption{Image of the qubit-resonator coupled system.
(A)~False-colored optical image of the device magnified at the qubit part. The qubit (white) is coupled to the center conductor of the coplanar waveguide (CPW) resonator (green) through a capacitance of $4$~fF.
(B)~Scanning electron micrograph of the three-junction flux qubit. The areas shaded by yellow indicate the Josephson junctions.}
\label{FigS1}
\end{figure}

Our device is composed of a $\lambda /2$ superconducting coplanar waveguide (CPW) resonator and a superconducting flux qubit (Fig.~1A).
The CPW resonator is made of a 50-nm-thick Nb film sputtered on a 300-$\mu$m-thick undoped silicon wafer with a 300-nm-thick thermal oxide on the surface. 
It is patterned by electron-beam (EB) lithography using the ZEP520A-7 resist and ${\rm CF_4}$ reactive ion etching.
The flux qubit with three Josephson junctions, where one is made smaller than the other two by a factor of $\alpha$, is fabricated by EB lithography and double-angle evaporation of Al using PMMA ($50$~nm)/Ge ($50$~nm)/MMA ($400$~nm) trilayer resist (Fig.~\ref{FigS1}B).
The thicknesses of the bottom and the top Al layers separated by an ${\rm Al_2O_3}$ layer are 20 and 30 nm, respectively.
In order to realize a superconducting contact between Nb and Al, the surface of Nb is cleaned by Ar ion milling before the evaporation of Al.
The qubit is located at one end of the resonator and is coupled to the resonator dispersively through a capacitance of 4~fF, while it is coupled to the drive port inductively (Fig.~\ref{FigS1}A).

The flux qubit is always biased with a half flux quantum where the transition frequency of the qubit $\omega_{\rm ge}$ from the ground state $\vert g \rangle$ to the excited state $\vert e \rangle$ is $2\pi \times 5.508$~GHz ($T_1 \sim 700$~ns during photon-detection experiments), while the resonator frequency $\omega_{\rm r}$ is $2\pi \times 10.256$~GHz ($Q$ factor $\sim 630$) when the qubit is in the $\vert g \rangle$ state.
It is shifted by a dispersive interaction with the qubit of $-2\chi=-2\pi \times 69$~MHz, which is enhanced by the straddling effect and the capacitive coupling~\cite{Inomata12} when the qubit is in the $\vert e \rangle$ state.
Note that $\omega_{\rm ge}$ and $\omega_{\rm r}$ denote not their bare frequencies but the renormalized ones including the dispersive shifts~\cite{Inomata14}.

A parametric phase-locked oscillator (PPLO)~\cite{Lin14}, which is previously operated as a flux-driven Josephson parametric amplifier (JPA)~\cite{Yamamoto08} consists of a $\lambda /4$ superconducting CPW resonator terminated by a dc-SQUID (superconducting quantum interference device).
A pump port is coupled to the SQUID loop inductively.
The device was fabricated by the planarized Nb trilayer process at MIT Lincoln Laboratory.
The resonator and the pump port are made out of a 150-nm-thick Nb film sputtered on a Si substrate covered by a 500-nm-thick SiO$_2$ layer.
A static resonant frequency of the PPLO is $\omega_{\rm r}^{\rm PO}=2\pi \times 10.948$~GHz.
The PPLO chip is the same as the one used in Ref.~\cite{Lin14}.

\subsection{Experimental setup}
\begin{figure*}[]
\includegraphics[width=12.4cm]{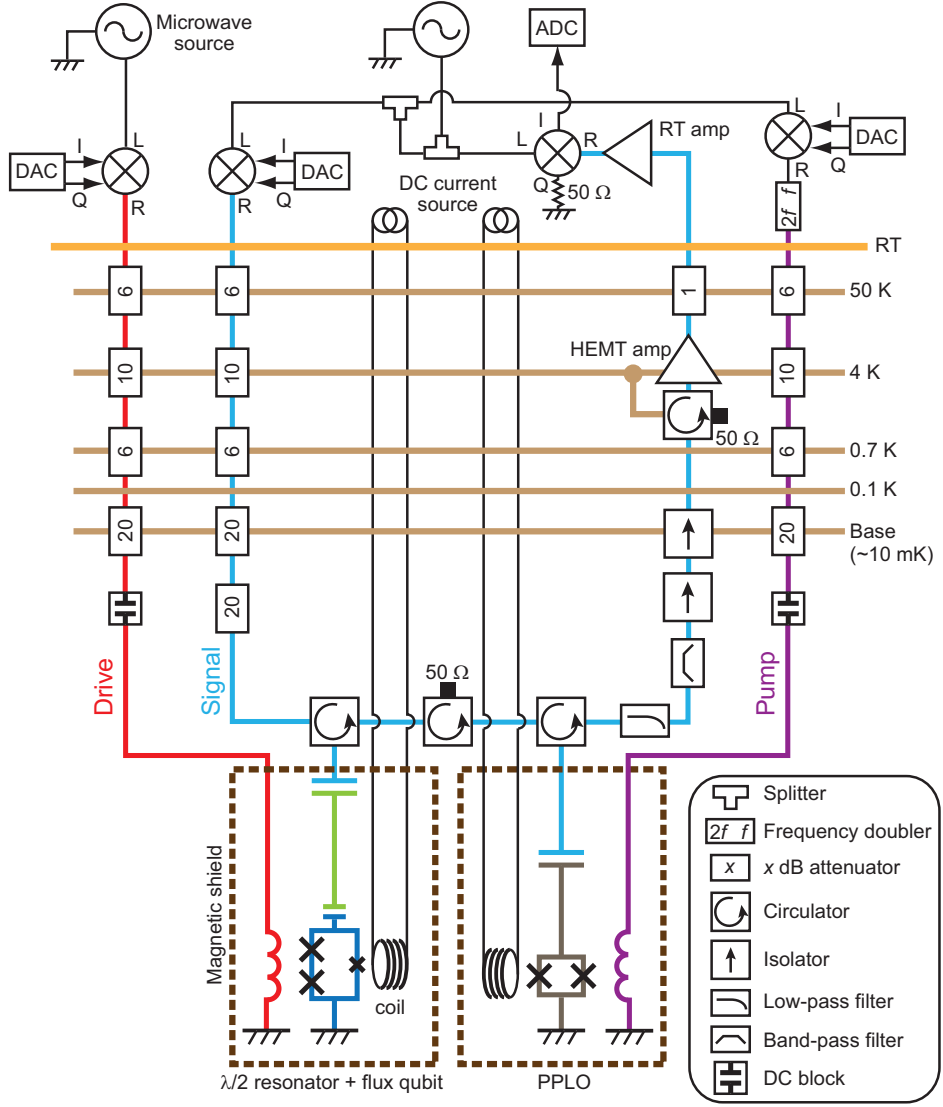}
\centering
\caption{Experimental setup diagram.}
\label{FigS2}
\end{figure*}

A schematic of the measurement setup including the wiring in a cryogen-free $^3$He/$^4$He dilution refrigerator, circuit components, and instruments used in the experiment is shown in Fig.~\ref{FigS2}.

The qubit$+$resonator and the PPLO circuits are fabricated on separate chips and are separately mounted in microwave-tight packages equipped with an independent coil for the flux bias.
They are protected independently by the Cryoperm magnetic shield from an external flux noise such as the geomagnetic field.

Microwave pulses for the drive, signal, and pump ports are generated by mixing the continuous microwaves with pulses which have independent IF frequencies generated by DACs (digital to analog converter) developed by Martinis group at UCSB~\cite{DACNote}.
The pulses are applied through the input microwave semi-rigid cables, each with attenuators of $42$ dB in total, and DC-blocks for the drive and pump ports.
For the signal port, the microwave pulses are further attenuated by 20~dB, and are input to the resonator through a circulator to separate the input and reflected waves.
The reflected waves are routed to PPLO via three circulators ($9$-$11$~GHz) and are reflected there again, and are propagated through a low-pass ($f_c=12.4$~GHz) and band-pass filters ($9$-$11$~GHz), two isolators ($9$-$11$~GHz), and the circulator ($9$-$11$~GHz) with a $50~\Omega$ termination.
Finally, the signals are amplified by a cryogenic HEMT amplifier and a room-temperature amplifier with a total gain of $\sim 66$~dB, and mixed with a local oscillator at an I/Q mixer down to the IF frequency.
The I component of the reflected signals are sampled at 1~GHz/s by an ADC (analog to digital converter).

For the impedance-matching measurement (Fig.~2A), the PPLO is kept off.
Namely, pump pulse is off (the output from the DAC in the pump port is turned off) and $\omega_{\rm r}^{\rm PO}$ is far detuned from $\omega_{\rm r}$ so that the PPLO acts as a perfect mirror.
In other measurements, the PPLO is kept on.

\subsection{Input-power calibration}
\begin{figure}
\includegraphics[width=5.7cm]{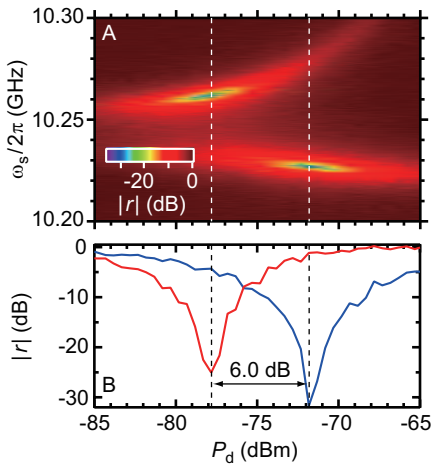}
\centering
\caption{(A)~Amplitude of the reflection coefficient $\vert r \vert$ of a continuous input signal as a function of its frequency $\omega_{\rm s}$ and the drive power $P_{\rm d}$. Two dips corresponding to absorptions of the input microwave due to the impedance matching are observed.
(B)~Cross-sections of A at $\omega_{\rm s}/2\pi = 10.227$~GHz (blue curve) and $10.262$~GHz (red curve). Difference in $P_{\rm d}$ between two dips is $P_{\rm diff} = 6.0$~dB.}
\label{FigS3}
\end{figure}

To estimate the photon detection efficiency precisely, calibration of the signal microwave power level on the sample chip is required.
For the calibration, we measure the reflection coefficient as a function of the signal microwave frequency $\omega_{\rm s}$ and the drive power $P_{\rm d}$ and determine the impedance-matching points (Fig.~\ref{FigS3}).
Here, we use continuous microwaves for both the signal and the qubit drive, and set the drive frequency at $\omega_{\rm d} = \omega_{\rm ge} -2\pi \times 46$~MHz.
We observe two dips representing the impedance matching, similarly to the inset of Fig.~2A. 
In the limit of weak signal power and no intrinsic loss of the resonator, these dips are expected to appear at the same $P_{\rm d}$, where the two radiative decay rates of the $\Lambda$ system are balanced, $\tilde{\kappa}_{31} = \tilde{\kappa}_{32}$ and $\tilde{\kappa}_{41} = \tilde{\kappa}_{42}$~\cite{Koshino13}, where $\tilde{\kappa}_{\rm ij}$ is the radiative decay rate for the $\vert \tilde{i} \rangle \rightarrow \vert \tilde{j} \rangle$ transition in the impedance-matched $\Lambda$ system.
In the actual system, however, the finite population in the level $\vert \tilde{2} \rangle$ as well as the intrinsic loss of the resonator weakens the elastic photon scattering from the $\Lambda$ system, and the impedance matching occurs when the radiative decay rates are not balanced, 
$\tilde{\kappa}_{31} > \tilde{\kappa}_{32}$ and $\tilde{\kappa}_{41} > \tilde{\kappa}_{42}$~\cite{Inomata14}.
This yields a difference in the drive power, $P_{\rm diff}$, between the two dips. 
$P_{\rm diff}$ is sensitive to the input signal power: As we increase the signal power, the level $\vert \tilde{2} \rangle$ is more populated and $P_{\rm diff}$ gets larger. Note that the small $P_{\rm diff}$ observed in the inset of Fig.~2A is attributed to the intrinsic loss of the resonator, since the pulsed signal field is sufficiently weak in this measurement.

We use $P_{\rm diff}$ to calibrate the signal power level.
We determine the signal power level which reproduces $P_{\rm diff} = 6.0$~dB (Fig.~\ref{FigS3}B) by the numerical simulation, following Ref.~\cite{Koshino13}.
In the numerical simulation, we employ the following parameters which are estimated by independent measurements: the qubit decay rate $\Gamma /2\pi = 0.174 \pm 0.012$~MHz (during this measurement $T_1$ shows $\Gamma^{-1} = 919 \pm 62$~ns) and the ratio of the external and total decay rates of the resonator photon $\kappa_{\rm ext}/\kappa = 0.964 \pm 0.003$ (for other parameters, see ``Device" section of this supplementary material).
As a result, the signal power is estimated to be $P_{\rm s} = -145.28$~dBm at
maximum ($\Gamma /2\pi = 0.186$~MHz and $\kappa_{\rm ext}/\kappa = 0.967$) and $P_{\rm s} = -146.02$~dBm at minimum ($\Gamma /2\pi = 0.162$~MHz and $\kappa_{\rm ext}/\kappa = 0.961$).
Therefore, $P_{\rm s} = -145.65 \pm 0.37$~dBm. 
This agrees well with an independent estimation of $P_{\rm s} = -146.0$~dBm by taking into account the total losses in the input port.

\subsection{Protocol for single photon detection}
In the main text (Fig.~1C), we show the protocol for single photon detection. 
Here, we present the detailed parameters of the pulses in the protocol.

The drive frequency is set at $\omega_{\rm d}=\omega_{\rm ge} - \delta \omega$, where $\delta \omega=2\pi \times 49$~MHz ($<2\chi$) is the detuning from the qubit energy, and is fixed through all the experiments described in the main text.
The drive pulse is synchronized with the signal pulse with a Gaussian envelope with a length $t_{\rm s}$ corresponding to its full width at half maximum (FWHM) in the voltage amplitude. 
The duration $t_{\rm d}$ of the drive pulse is optimized as $t_{\rm d} = 1.5 t_{\rm s} + 50$~ns so that the signal pulse is completely covered by the drive pulse and is efficiently absorbed by the $\Lambda$ system.
In order to suppress unwanted nonadiabatic qubit excitations, the rising and falling edges of the drive-pulse envelope are smoothed by Gaussian function with FWHM of $2t_{\rm rise}=30$~ns in the voltage amplitude.

The readout pulse (the frequency $\omega_{\rm rd}=\omega_{\rm r}-2\chi=2\pi \times 10.187$~GHz, the length $t_{\rm rd} = 60$~ns, and the mean photon number $\bar{n}_{\rm rd} \sim 10$) is applied after the delay of $t_{\rm delay1}=t_{\rm d}/2 + t_{\rm rise}$ from the center of the drive and signal pulses.
The reflected readout pulse works as a locking signal for the PPLO output phase, and the pump pulse (the frequency $\omega_{\rm pump} = 2\omega_{\rm rd}$, the length $t_{\rm pump} = 400$~ns, and the power $P_{\rm pump} \sim -60$~dBm) is applied after $t_{\rm delay2} = 40$~ns.
The parametric oscillation signal with either $0$ or $\pi$ phase is output from the PPLO during the application of the pump pulse, and the data acquisition time of $\sim 100$~ns is required to extract the phase.

\subsection{Photon detection efficiency}
In Figs.~2B, 2E, and 3A, a mean photon number in a signal pulse $\bar{n}_{\rm s}$ is kept to be $\sim 0.1$, which implies that $\sim 5\%$ of the weak-coherent signal pulses contain multiple photons. 
Our detector responds to the multi-photon pulses but cannot discriminate them from single-photon pulses.
The efficiency $\eta$ includes those counts.

\subsection{Dark count in the detector}
\begin{figure}[]
\includegraphics[width=5.6cm]{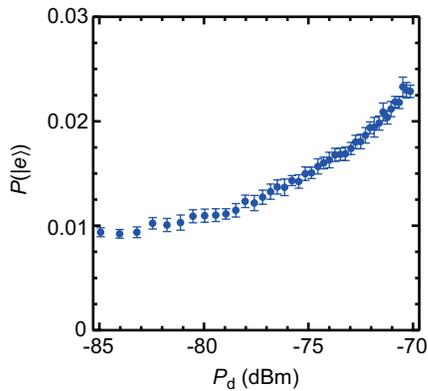}
\centering
\caption{Dark count probability in the detector. The data was taken ten times each and averaged before and after the measurement in Fig.~2B. The dark count probability including the imperfect initialization shows $0.014 \pm~0.001$ at $P_{\rm d}=-75.5$~dBm where the single-photon-detection efficiency hits the maximum. The error bars represent the standard deviation in twenty identical measurements.}
\label{FigS4}
\end{figure}

Figure~\ref{FigS4} shows the dark count probability in the detector, which is the click probability without applying the signal pulse in the pulse sequence of Fig.~1C.
The dark count is mainly caused by the nonadiabatic qubit excitation due to the drive pulse and the imperfect initialization.
The probability induced by the latter factor is constant and is measured to be $0.008 \pm 0.001$, while the probability induced by the former factor depends on the power and the length of the drive pulse and remains finite even with the Gaussian envelope.
We determine the dark count probability before and after each measurement of Figs.~2B,~2E, and Fig.~3 and subtracted the averaged value from the measurement result.

\subsection{Time constant of an impedance-matched $\Lambda$ system}
\begin{figure}[]
\includegraphics[width=5.6cm]{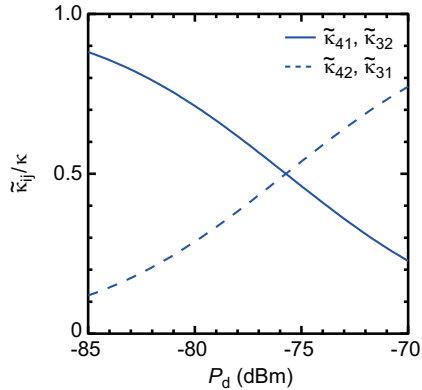}
\centering
\caption{Radiative decay rates of the impedance-matched $\Lambda$ system, which are calculated based on the experimental parameters, as a function of the drive power. The two relevant decay rates, $\tilde{\kappa}_{41}$ and $\tilde{\kappa}_{42}$ or $\tilde{\kappa}_{31}$ and $\tilde{\kappa}_{32}$, become identical at $P_{\rm d} = -75.7$~dBm, where the impedance matching takes place.}
\label{FigS5}
\end{figure}

We denote the overall decay rate of the resonator by $\kappa$ and the radiative decay rate for the $\vert \tilde{i} \rangle \rightarrow \vert \tilde{j} \rangle$ transition in the $\Lambda$ system by $\tilde{\kappa}_{\rm ij}$.
Figure~\ref{FigS5} shows $\tilde{\kappa}_{\rm ij}/\kappa$ as a function of the drive power $P_{\rm d}$, calculated based on the experimental parameters. 
In the experiment, we choose $P_{\rm d} = -75.5$~dBm where the photon detection efficiency $\eta$ reaches the maximum.
At this point, $\tilde{\kappa}_{\rm 41}/\kappa = 0.49$.
The time constant of the impedance-matched $\Lambda$ system for the voltage amplitude decay, $\tau_{\Lambda}$, is estimated to be $2/\kappa \sim 20$~ns, where $\kappa = \tilde{\kappa}_{\rm 41} + \tilde{\kappa}_{\rm 42} \sim 2\pi \times 16$~MHz.
The shortest signal pulse length is $34$~ns in Fig.~3B, which is comparable with $\tau_{\Lambda}$.

\subsection{Protocol for reset}
In the main text (Fig.~4A), we show the reset protocol for the single photon detection. Here, we describe how to optimize the parameters of pulses in the protocol.

We first apply a $\pi$ pulse with the length of $6$~ns to directly excite the qubit from the $\vert g, 0\rangle$ to the $\vert e, 0\rangle$ state.
Then, we apply the drive and reset pulses to induce the $\vert \tilde{2} \rangle \rightarrow \vert \tilde{3} \rangle \rightarrow \vert \tilde{1} \rangle$ transition.
To find the operating point to maximize the resetting efficiency, we swept the frequency $\omega_{\rm rst}$ of the reset pulse and the drive power $P_{\rm dr}$.
After fixing $\omega_{\rm rst}$ and $P_{\rm dr}$, we optimize the drive pulse length $t_{\rm dr}$, and the mean photon number in the reset pulse $\bar{n}_{\rm rst}$ to minimize $P(\vert e \rangle)$.
Finally, we measure $P(\vert e \rangle)$ as a function of $\omega_{\rm rst}$ and $P_{\rm dr}$ using the reset protocol with optimized parameters.
Parameters for the readout and pump pulses are the same as the ones in Fig.~1C.

At the optimal reset point, $P(\vert e \rangle)$ shows the minimum value of $0.017 \pm 0.002$, which results in twice larger occupation of the qubit excited state compared to $0.008 \pm 0.001$ obtained in the initialization in the equilibrium condition. This indicates the small probability of unwanted nonadiabatic excitations by the drive pulse in the reset protocol.

We demonstrate the microwave photon detection combined with the fast reset protocol. 
We apply the drive and the signal pulses (the same conditions as in the measurement in Fig.~2B) after the reset protocol and readout the qubit. 
We accomplish $\eta$ of $0.67 \pm 0.06$ which is consistent with the maximum $\eta$ obtained in Fig.~2E. 
This indicates that the photon detection efficiency is unaffected by the reset protocol.

It takes $410$~ns to reset the system and $208$~ns to detect the single photon for $t_{\rm s} = 85$~ns.
Both of the durations are determined by the drive pulse widths including $t_{\rm rise} = 15$~ns.
The qubit readout is completed by accumulating data for $100$~ns after $t_{\rm delay2} = 40$~ns.
The period of the single photon detection including the reset protocol is $\sim 760$~ns, which allows the photon counting rate of $\sim 1.3$~MHz.



\end{document}